\documentclass[%
 reprint,
 amsmath,amssymb,
 aps,
]{revtex4-1}

\usepackage{graphicx}
\usepackage{dcolumn}
\usepackage{bm}
\usepackage{color}

\def\Ni{{{\mathcal N}_i}}

\begin{document}


\title{Differentiating swarming models by mimicking a frustrated anti-Ferromagnet}

\author{D. J. G. Pearce${}^{1,2}$}
\email{d.j.g.pearce@warwick.ac.uk}

\author{M. S. Turner${}^{1,3}$}
 \email{m.s.turner@warwick.ac.uk}
\affiliation{University of Warwick Department of Physics${}^{1}$, MOAC Doctoral training centre${}^{2}$ and Centre for Complexity Science${}^{3}$, Coventry, CV4 7AL, United Kingdom}

\date{\today}

\begin{abstract}

Self propelled particle (SPP) models are often compared with animal swarms. However, the collective behaviour observed in experiments usually leaves considerable unconstrained freedom in the structure of these models. To tackle this degeneracy, and better distinguish between candidate models, we study swarms of SPPs circulating in channels (like spins) where we permit information to pass through windows between neighbouring channels. Co-alignment between particles then couples the channels (antiferromagnetically) so that they tend to counter-rotate. We study channels arranged to mimic a geometrically frustrated antiferromagnet and show how the effects of this frustration allow us to better distinguish between SPP models. Similar experiments could therefore improve our understanding of collective motion in animals. Finally we discuss how the spin analogy can be exploited to construct universal logic gates and therefore swarming systems that can function as Turing machines.
\end{abstract}

\maketitle

Collective motion in large groups of animals represents one of the most conspicuous displays of emergent order in nature \cite{vortexshoalimage,Murm,Bats,pedestrian}.  The idea that such swarms manifest some kind of effective group intelligence has been explored in several recent studies \cite{couzinNature,sumpter2008consensus,conradtNature,couzin2005effective}. While swarming is ubiquitous in nature it is still surprisingly poorly understood. In particular there is a large space of candidate agent-based models, some of which have been studied in detail \cite{giardinarev,vicsek,vicsekCOM,swarmmodel2,chatevoronoi,OrsognaAR,HHmodel,HHmodel2012,couzinAR,chateAR,ARmodelreview}. Typically, a rule for the motion of every individual is first specified and the resulting collective motion is then studied. However, it can be very difficult to refine this ``microscopic" rule by studying data for the collective ``macroscopic" behavior. The essential difficulty is that model building is an inverse problem in which no techniques yet exist to perform this inversion.

Recent experiments have studied animal swarms in confined environments \cite{locustrings,ward2008quorum,fishinaring,sumpter2008information} even managing to create behavior that mimics logical operations \cite{li2011bacteriaAND,gunji2011crabOR}. Our work was primarily motivated by one such experiment performed on locusts enclosed in a single ring-shaped channel \cite{locustrings} where increasing the density of locusts results in a transition from a state of random motion to a polarised state in which the locusts co-align to create coherent, circulating swarms. Due to the ring-shaped enclosure the swarm was able to polarise into clockwise or anticlockwise circulation, giving it a spin-like nature. 
\begin{figure}[htp]
\includegraphics[width = \columnwidth]{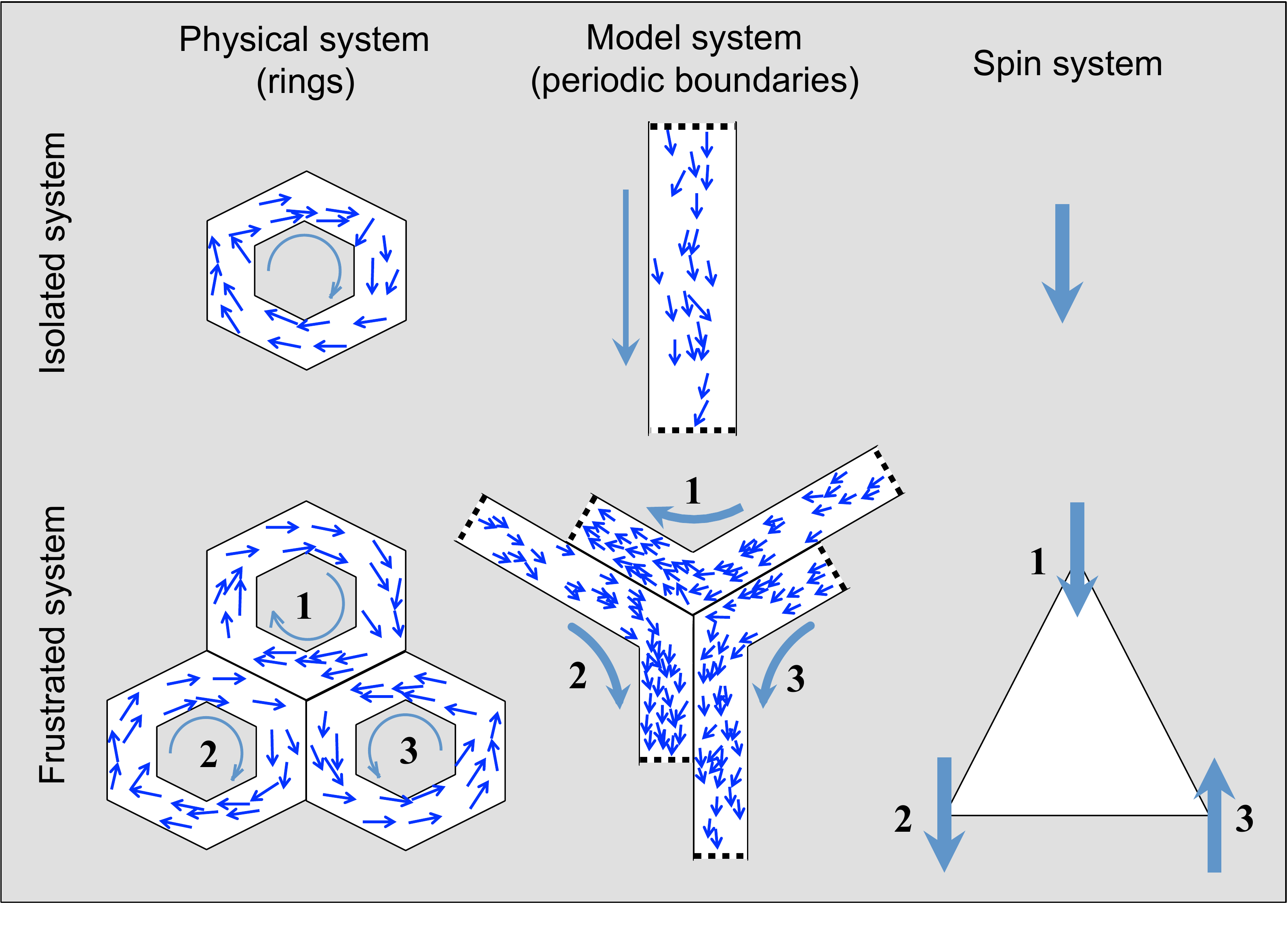}
\caption{\label{fig:rings} Different self-propelled particle (SPP) models are studied in confining channels. {\bf Isolated system}: The macroscopic behavior of a ring containing swarming animals is approximated by interacting agents moving in a linear, semi-periodic channel, for simplicity. Clockwise/anticlockwise collective motion in the ring, analogous to a spin, corresponds to motion up/down the semi-periodic channel. {\bf Frustrated system}: The motion within three rings arranged on a triangular lattice is frustrated when interactions are permitted across windows between the tracks. This is again simulated using linear semi-periodic channels for the SPP model (which remain linear but are shown as kinked in the middle panel for clarity; periodic linear channels with windows between all pairs cannot easily be represented in a 2D image). This system is analogous to a geometrically frustrated anti-ferromagnet.}
\end{figure}
This behavior was compared with a simple one dimensional SPP model with periodic boundary conditions, see the {\it isolated system} panels in Fig.~\ref{fig:rings}. The polarization transition and the mean time between spontaneous polarization inversions was then related to the parameters of the model \cite{locustrings}. However, we believe that it is hard to draw any definite conclusions concerning the correct {\it structure} for the model as there remains considerable freedom to choose structurally and parametrically distinct SPP models that would all be capable of reproducing this stylized behavior. It is a challenging task to distinguish structurally distinct candidate models by comparison with data like this. Our approach is to seek to break the behavioral degeneracy between models. In order to achieve this we first consider two ring-shaped channels arranged near to one-another that share a (section of) boundary through which the individuals can pass information but cannot physically cross. This could be realised experimentally by connecting the rings by a window. In animals that mainly employ a  sense of vision a transparent window would be appropriate; for animals that use touch a limited physical opening might be used. This window provides a coupling between the two rings. Here we extend the interactions between individuals to include neighbours that are {\it visible} through the window, as well as those that are visible within the same ring, and use the same behavioral rule for both cases. For highly polarised swarms, driven by co-alignment, we would then expect a ring polarised anticlockwise (an ``up" spin) to be most stable when it is adjacent to a ring polarised clockwise (a ``down" spin), or vice-versa: Only in this situation would neighbours connected through the window also find themselves co-aligned. The coupling across the window is therefore {antiferromagnetic} in character. 

Inspired by the extensive literature on frustrated antiferromagnetic systems \cite{andreanov2010spin,reimers1992absence,ramirez1994strongly} we analyse motion in three rings arranged so they each each share a boundary with the other two, see the {\it frustrated system} panels Fig.~\ref{fig:rings}. In this way we create a system similar to geometrically frustrated antiferromagantic atoms on a triangular lattice. It is no longer possible for all three rings to remain highly polarised and co-aligned across all windows. As in the analogous magnetic system we no longer expect a unique pair of symmetric ground states to exist. We anticipate that additional information can be obtained from the resulting behavior, whatever it may be, that can be used to better distinguish microscopic models when they are constrained against observed behavior.

In what follows we compare two different SPP models frustrated in this way.  Apart from the boundary conditions both take a fairly standard form in which $N$ particles move in a periodic box with a constant speed $v_0=1$. When combined with a (unit) time step this defines our units of length throughout. At each discrete time step every particle orientates its velocity along the average direction of motion of its neighbours. The only difference between the two models studied here will be how these neighbours are identified. Writing those neighbours to the $i^{\rm th}$ particle as $\Ni$ the equation of motion involves the average velocity of its neighbours $\widehat{\langle{\underline{v}} ^{t}_{j} \rangle}_{j\in \Ni}\equiv\frac{\sum_{j\in\Ni}{\underline{v}} ^{t}_{j}}{|\sum_{j\in\Ni}{\underline{v}} ^{t}_{j}|}$. Noise is introduced by randomly orientated unit vectors $\hat{\underline{ \eta }}^{t}_{i}$ that are uncorrelated between individuals and in time $\langle \hat{\underline{ \eta }}^{t}_{i}\cdot \hat{\underline{ \eta }}^{t'}_{j}\rangle=\delta_{ij}\delta_{tt'}$. The position, $\underline{r}^t_{i}$, and velocity, $\underline{v}^t_{i}$, of particle, $i$ at time $t$ are then given by the following equations, where the parameter $\phi_{n} < 1$ controls the relative weighting of the noise term and a hat $\hat{}$ indicates a unit vector throughout.
\begin{eqnarray}
\underline{v}^{t+1}_{i}= (1-\phi_{n})\widehat{\langle{\underline{v}} ^{t}_{j} \rangle}_{j\in \Ni} +\phi_{n}\hat{\underline{ \eta }}^{t}_{i}
\label{eq:v}
\\
\underline{r}^{t+1}_{i}= \underline{r}^{t}_{i}+v_{0}\hat{\underline{v}}^{t}_{i}
\label{eq:r}
\end{eqnarray}
The first of our models is typical of a class that identify nearest neighbours according to a {\bf metric}-based measure of distance (the model due to Vicsek and coworkers \cite{vicsek} is often cited as a prototype). Here a particle co-aligns with others that lie within a fixed interaction range $R$. This definition means that individuals can have as few as zero or as many as $N-1$ neighbours. The second model selects nearest neighbours according to a {\bf metric-free} scheme, motivated by the evidence for interactions with this character in bird flocks \cite{ScaleFree,ballerinitopol}. In this model each particle aligns with the $N_c$ nearest particles, irrespective of absolute separation. Both these models generally exhibit two distinct states, {\it ordered}, in which the particles achieve a high level of polarization and all their velocities are locally highly aligned, and {\it disordered}, in which there is no net polarization and the velocities of individuals are largely uncorrelated. The transition from the disordered to ordered state is primarily controlled by two quantities: the noise weighting $\phi_{n}$ and the density of particles. For sufficiently low noise, $\phi_{n}$, and high density the system is ordered. As the noise is increased (or the density is decreased) the system undergoes a transition into the disordered state. Here, we simulate swarms of $N=100$ SPPs in a semi-periodic box of width  and height $W=H=2.5$ and length $L=25$ in the $x$, $y$ and $z$ direction respectively. This is an unconventional choice in that the system is only periodic in the $z$ direction, instead of in $x$, $y$ and $z$. If a particle reaches a boundary perpendicular to the $x$ or $y$ directions it undergoes an elastic collision, or reflection, in which the component of its velocity perpendicular to that boundary is reversed. In this way the swarm can be confined to a slender, periodic channel, see SI for details. This leaves three free control parameters, the number of particles, $N$, the noise weighting, $\phi_n$, and the interaction range, $R$ for the metric and $N_c$ for the metric-free models (see SI for details). Simulation were computed for 90,000 time steps after a 10,000 time step pre-equilibration (except for the spatial inhomogeneity data where ten realisations each of 10,000 time steps were more appropriate).

Due to the nature of the semi-periodic box, the swarm cannot sustain a high level of polarization unless it is aligned nearly parallel, or anti-parallel, to the $z$ axis. This is because orientation in either the $x$ or $y$ direction will result in collisions with the non-periodic boundaries and the individuals in the swarm will then rapidly change direction in an incoherent fashion until order along $z$ re-emerges. For this reason it is possible to quantify the polarization of the system using only the $z$-component of velocity, analogous to the polarization of circulation.
\begin{equation}
P_{z}^{t} = \frac1N \sum \underline{v}_{i}^{t}\cdot \hat{\underline{z}}
\label{eq:mu}
\end{equation}
For disordered swarms $P_{z}^{t} \approx 0$, and for highly ordered swarms $P_{z}^{t} \approx \pm 1$. 

\begin{figure}[htp]
\includegraphics[width = \columnwidth]{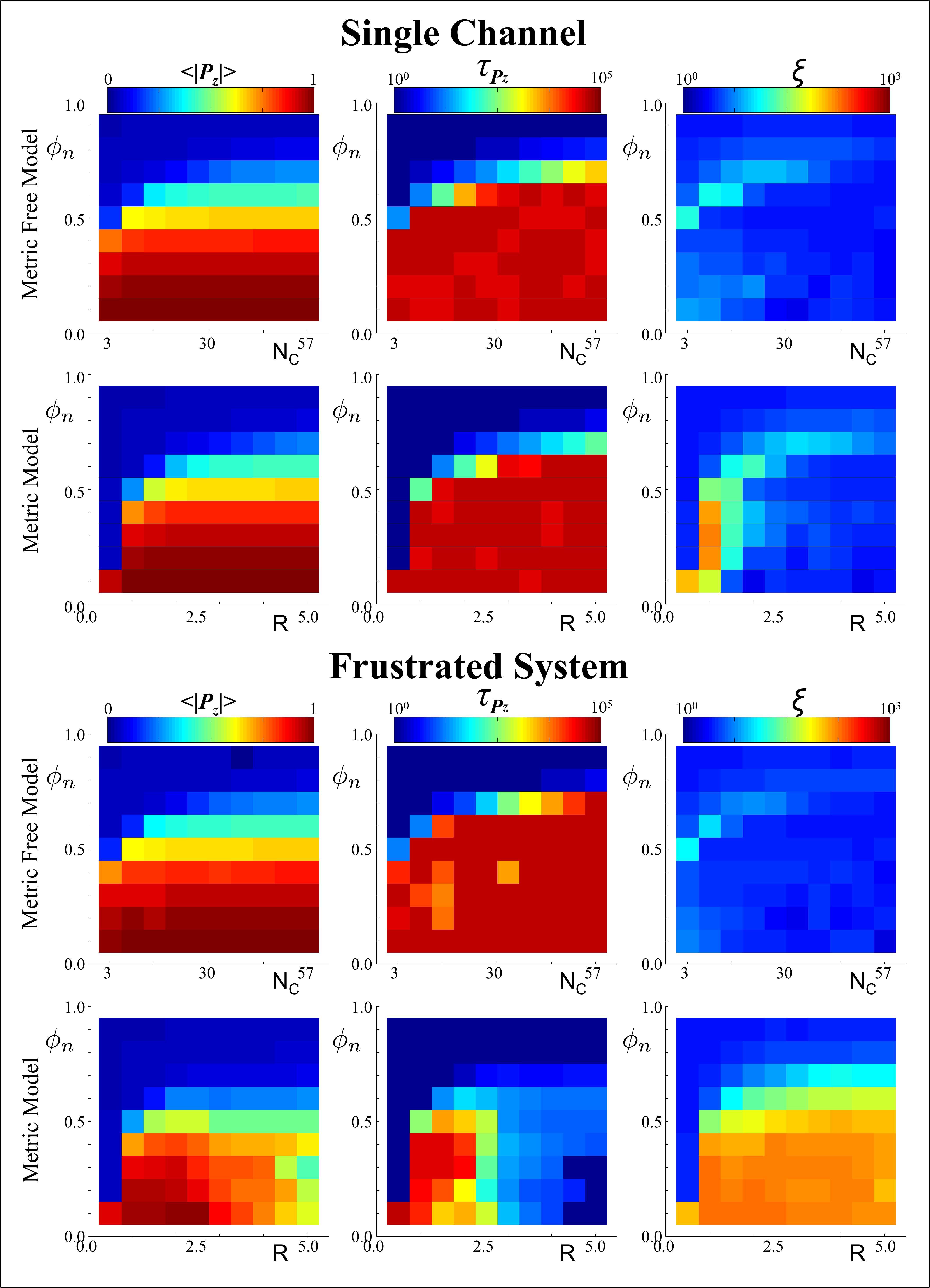}
\caption{\label{fig:Vphase} The behavior of the metric and metric-free models can be more easily distinguished when the system is frustrated. Simulations are performed in a single channel (top panel), and a system of three fully frustrated channels (bottom panel). Shown is the average polarization magnitude ($\langle|P_z|\rangle$, left column), polarization correlation, or persistence, time ($\tau_{Pz}$, middle column), and the spatial inhomogeneity ($\xi$, right column) for various interaction ranges ($R$ or $N_c$ for metric or metric-free models respectively) and noise levels ($\phi_n$), see text for details.}
\end{figure}

Swarms of SPPs confined in these channels support both ordered and disordered phases (with high and low polarizations, respectively), with a transition between the two around $\phi_n \sim 0.5$, see Fig.~\ref{fig:Vphase} (single channel). Near this transition the swarms are polarised, $P_{z}^{t} \sim 0.5$ and have a clear direction of motion along the channel, but there is still sufficient noise that the swarm can reverse direction, evidenced by the autocorrelation times for $P_{z}^{t}$. As $\phi_n$ is decreased, the rate of these directional switches decreases and the direction of polarization eventually no longer changes on  timescales that are accessible in our simulations. A similar outcome is observed for both SPP models, reproducing the behavior of insect swarms enclosed in a ring and previous simulations thereof \cite{locustrings}.

In order to introduce a coupling between two adjacent channels they are positioned alongside each other so that they share a face normal to the $y$ axis (say), i.e. particles in channel 1 can be thought of as being restricted to $x \in [0,W]$ and particles in channel 2 to $x \in [-W,0]$. This means that the minimum distance between two particles in different channels is zero and particles in different channels can co-align if the line-of-sight connecting them passes through a region designated as a window. No transport of particles is allowed across the window. We can adjust the degree of coupling between two channels by changing the length of the windows. With pairwise coupling between three channels we can arrange them so as to be mutually frustrating, see Fig.~\ref{fig:rings}.

Since we are not restricted by geometrical considerations in these simulations, it is possible to extend the windows to run along the full length of the channel, with each channel sharing such a window with each of the other channels. We refer to this as a fully frustrated system. The results of simulations on such fully frustrated systems are as follows (see SI for details): For weak interactions (short ranges $R$ or small number $N_c$) little difference is observed in the behavior of the swarms, see Fig.~\ref{fig:Vphase} (frustrated system). However, when the interaction becomes stronger SPPs with metric-based interactions no longer reach long lived states with nearly constant polarization. When the interaction range becomes comparable to the width of the channels, $W \sim R$, two swarms in adjacent channels are unable to pass by each other without interacting. This often results in one of them reversing direction. It also acts to push the swarms into high density bands since the leading front  is the first to be affected by a band in another channel (see SI movies). This is evidenced by the higher values of $\xi$, defined as the maximum time-averaged variance in the number of particles observed in any constant fraction of the channel length. In contrast to this, SPPs with metric-free interactions exhibit high polarization and long polarization autocorrelation times, $\tau_{Pz}$. As these swarms clump into bands the majority of nearest neighbours remain sited in the same channel, which leads to a weaker coupling between swarms in adjacent channels; this allows them to pass each other without a significant effect on the polarization. Hence the fall in persistence times is not seen for metric-free swarms. We also studied partially frustrated systems in which the windows extend over only a third of their length (resembling the physical system sketched in the bottom left panel of Fig 1). This show qualitatively similar, but slightly weaker, effects (see SI).

In the final part of this letter we explore further the the spin-like nature of the motion within these channels to construct an information processing device. First we define a ``bit'' as 
\begin{equation}
m =\begin{cases}1 &\mbox{if } P_{z} > 0\\0 &\mbox{if } P_{z} < 0\end{cases} 
\label{eq:m}
\end{equation}

\begin{figure}[h]
\includegraphics[width = \columnwidth]{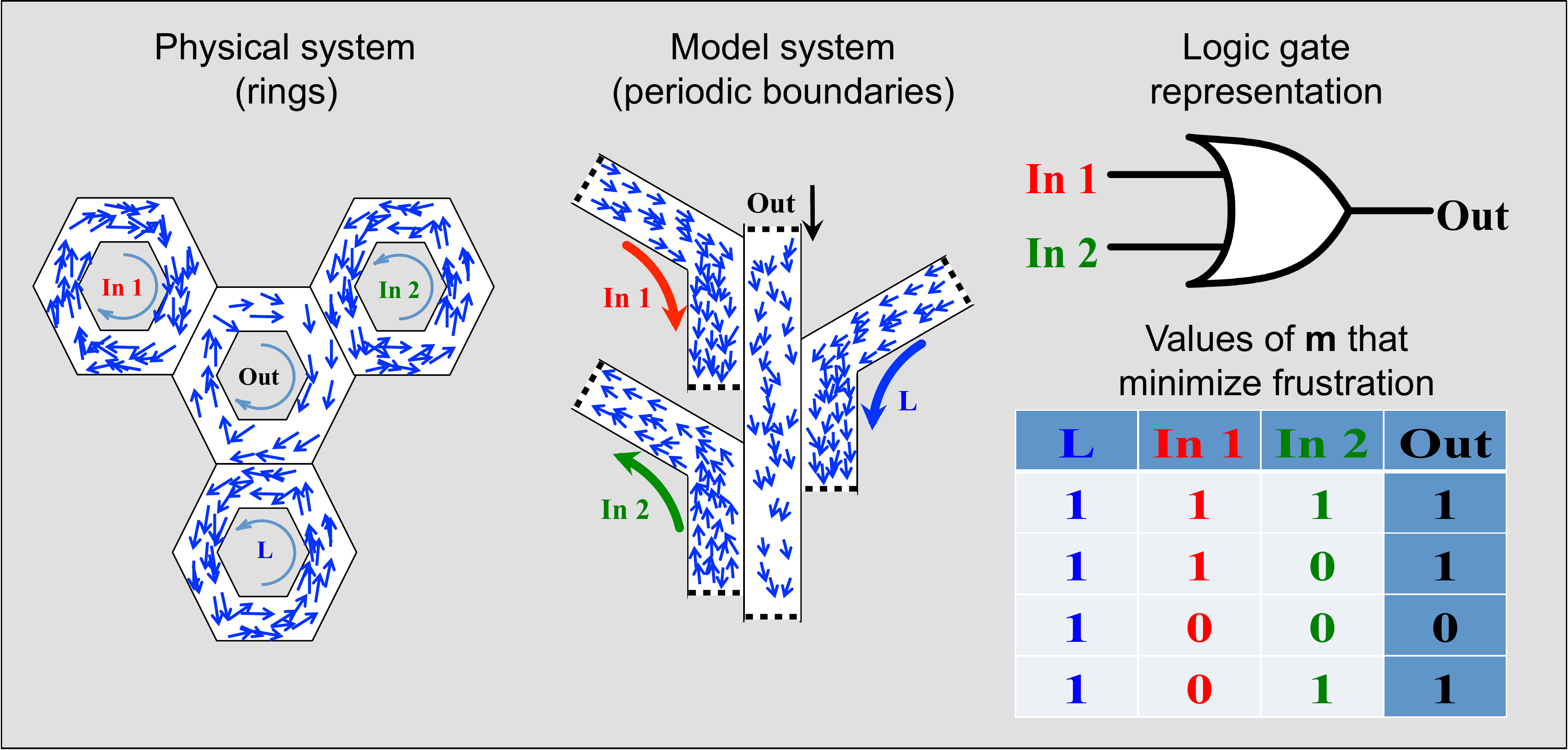}
\caption{\label{fig:logicdiag} Rings containing SPPs, as model for animal systems, that are predicted to perform a logical OR, with the direction of polarization of the \textbf{Out} ring the output, and \textbf{\textcolor{red}{In 1}} and \textbf{\textcolor{green}{In 2}} the inputs (left column). We study a SPP system in the corresponding arrangement of semi-periodic channels (middle column). The table shows the polarization directions that minimise frustration for different combinations of input polarizations (right column); this corresponds to the output of a logical OR gate.}
\end{figure}

We now study the arrangement of channels shown in Fig.~\ref{fig:logicdiag}. Here the polarization of the  \textbf{Out} channel depends on the polarizations of channels \textbf{In 1}, \textbf{In 2} and \textbf{L} (for {Locked}). Adopting a state that minimises the overall frustration would (and does) lead to the logic table shown in Fig.~\ref{fig:logicdiag}, which is equivalent to an OR gate \footnote{The NOT operation is trivial: a ring connected to the \textbf{Out} channel will have inverted spin due to the antiferromagnetic coupling, hence we also have access to universal NOR gates.}.

\begin{figure}[h]
\includegraphics[width = \columnwidth]{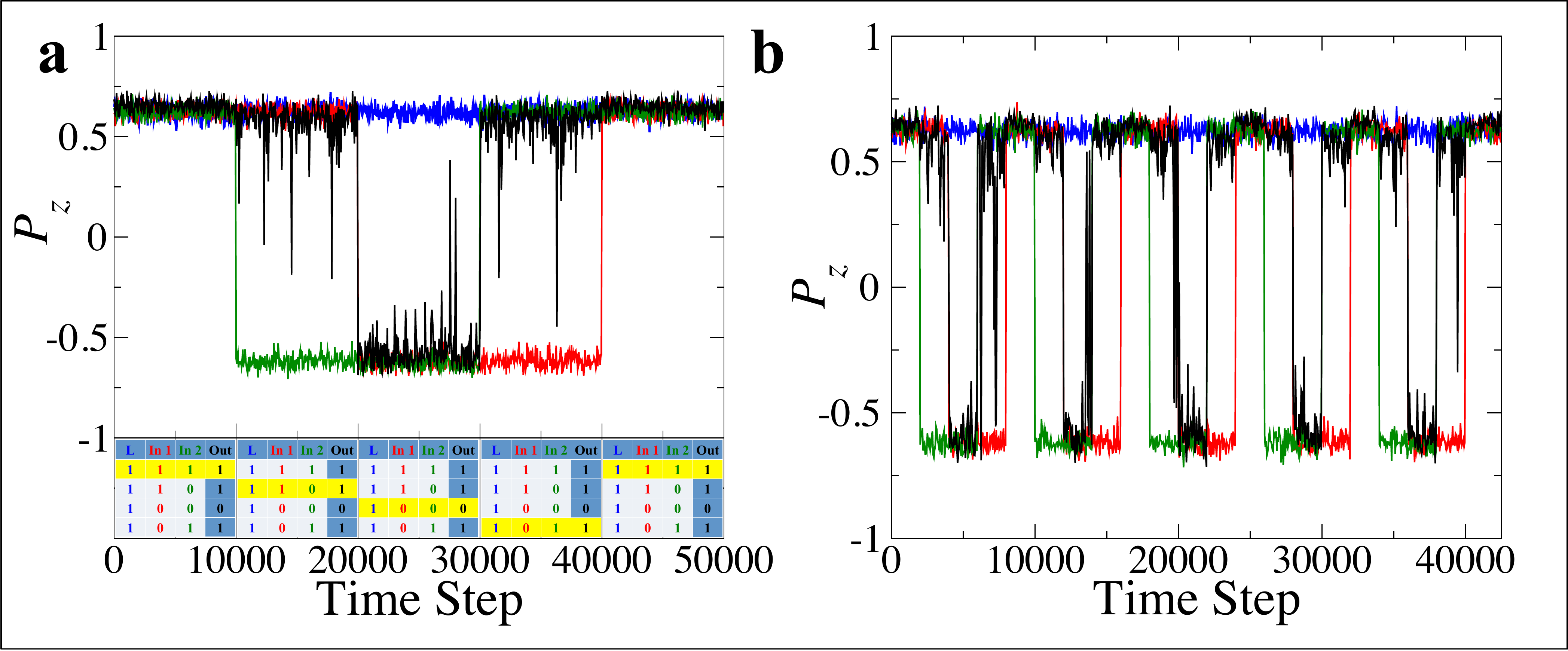}
\caption{\label{fig:logicOR} The polarization of SPP particles with metric-based interactions moving in the system shown in Fig.~\ref{fig:logicdiag}. The polarization in each channel is recorded over the course of a simulation run in which the particle polarizations in the \textbf{\textcolor{red}{In 1}} (red trace) and \textbf{\textcolor{green}{In 2}} (green trace) channels are manually inverted at intervals of (a) 10,000 and (b) 2,000 time steps. Also shown is the \textbf{Out} channel (black trace) and Locked channel (designated \textbf{\textcolor{blue}{L}}, blue trace), the latter maintaining a positive polarization through. After inversion the polarization of the \textbf{Out} channel undergoes a spontaneous and rapid transition to the state shown in the table under panel (a), if necessary, and will maintain this for more than $99\%$ of subsequent time steps under both slow (a) and fast (b) switching of inputs. Such systems can therefore mimic the behavior of a deterministic, rather than statistical, logic gates (see SI for details).}
\end{figure}

To validate that the logic table shown in Fig.~\ref{fig:logicdiag} is indeed realised we employ SPP particles with metric-based interactions of range $R=2.5$ and noise level $\phi_n=0.5$. In the absence of frustration this would lead to moderately polarised swarms with long persistence times, see Fig.~\ref{fig:Vphase}. We include fewer particles in the \textbf{Out} channel and assign it a width that is smaller than the interaction radius in order to make it reverse direction more rapidly than the \textbf{In} and \textbf{L} channels when it is frustrated (see SI for details). This results in an essentially deterministic logical output, rather than one that is only realised statistically, because the \textbf{Out} channel responds to the \textbf{In} channels, and not the other way around. This effectively means the \textbf{In} and \textbf{L} channels don't spontaneously reverse polarization. In order to probe the systems response we manually invert the directions of all particles in either of the \textbf{In} channels in order to study the corresponding outputs, see Fig.~\ref{fig:logicOR} and SI movies.

In summary, we show that different models can better be distinguished when the particle (animal) motion is frustrated. We achieved this by introducing windows through which particles confined to different channels can interact. We then use a channel geometry that mimics a geometrically frustrated antiferrognet. This approach promises to allow us to better distinguish between models for animal behaviour by comparing them with experimental data that is itself obtained in frustrated geometries. Ultimately this could lead to an improved insight into the behavioural mechanisms that lead to swarming, one of the prototypical examples of emergent order in nature.\\
\\

\vbox{Finally, we use a spin analogy to propose confining geometries in which the swarm(s) perform the operation of a universal logic gate. These could be combined to perform more complex computational tasks, placing a bound on the computational capability of animal swarms, at least those that are artificially confined in this way, to that of a Turing machine. Although the computation that is being performed in this class of confining geometries is unlikely to more than very loosely related to the computation that is being performed in swarms of unconfined animals our results underline the fact that there is no known limit to the emergent computational power of a swarm.}

\begin{acknowledgments}
This work was entirely supported by the UK Engineering and Physical Sciences Research Council through the MOAC Doctoral Training Centre (DJGP) and grant EP/E501311/1 (a Leadership Fellowship to MST). We also acknowledge use of the CGAL Library http://www.cgal.org in our simulations.
\end{acknowledgments}

%

\end{document}